\documentstyle[preprint,aps,epsfig,tighten,floats]{revtex}

\newcommand{\fdg}{\rlap{${}^{\hskip-0.5pt\circ}$}.}
\newcommand{\erf}{\mathop{\rm erf}}

\begin{document}

\title{
\null\vskip-0.9in{\small\hfill MCTP-01-31\\
\small\hfill CWRU-P15-01
\\\hfill June 2001}
\vskip0.3in
On the direct detection of extragalactic weakly interacting massive particles}

\author{Katherine Freese$^{1}$\thanks{Email address: ktfreese@umich.edu},
Paolo Gondolo$^{2}$\thanks{Email address: pxg26@po.cwru.edu},
Leo Stodolsky$^{3}$\thanks{Email address: les@mppmu.mpg.de}}

\address{$^1$Physics Department, University of Michigan, Ann Arbor, MI 48109,
  USA}

\address{$^2$Department of Physics, Case Western Reserve University, \\ 10900
  Euclid Avenue, Cleveland, OH 44106-7079, USA}

\address{$^3$Max Planck Institut f\"ur Physik, F\"ohringer Ring 6, 80805
  Munich, Germany}

\maketitle

\begin{abstract}
  We consider the direct detection of weakly interacting massive particles
  (WIMPs) reaching the Earth from outside the Milky Way. If these WIMPs form a
  distinct population they will, although of much lower flux than typical
  galactic halo WIMPs, have a number of features which might aid in their
  ultimate detectability: a high and essentially unique velocity ($\sim 600$
  km/s in the galactic rest frame) due to their acceleration in entering the
  Milky Way, and most likely one or two unique flight directions at the Earth.
  This high velocity may be experimentally advantageous in direct detection
  experiments, since it gives a recoil signal at relatively high energy where
  background is generally much reduced.  For a density of extragalactic WIMPs
  comparable to the critical density of the universe the count rate expected is
  very roughly the same as that of fast galactic WIMPs. If there is an
  increased density relative to critical density associated with the Local
  Group of galaxies, say 10-30 times the critical density, there is a
  corresponding increase in rate and the extragalactic WIMPs would show up as a
  high energy shoulder in the recoil energy distribution. Evidence of such
  WIMPs as a separate population with these distinct properties would offer
  interesting information on the formation and prehistory of the galaxy.
\end{abstract}

\section{INTRODUCTION}
\label{sec:intro}

The nature of the dark matter in the universe is one of the outstanding
questions in astrophysics and particle physics.  One of the favored candidates
is the WIMP (weakly interacting massive particle).  These particles are
presumed to have masses in the range 1 GeV/$c^2$ to a few TeV/$c^2$.

Much work has been done studying the possibilities for detecting these
particles.  Possibilities include direct detection \cite{goodman}, where the
particle interacts with a nucleus in a low temperature detector, and is
identified by the keV of energy it deposits in the nucleon; and indirect
detection, whereby the particles are captured in the Sun or Earth, sink to
their centers, annihilate with one another in the core, giving rise to
particles including neutrinos which can be detected by experiments on the
surface of the Earth \cite{Sun,Earth}. Alternatively, the particles
annihilate in the galactic halo~\cite{halo} or the galactic center~\cite{gon99}
and produce anomalous components in cosmic rays.

Most of this work has focused on detection of WIMPs that belong to the
virialized halo of the Milky Way galaxy. However, some authors
\cite{streams,copi,green,gelminigondolo,stiff} have started to consider the
detection of WIMPs that are not yet virialized with the Milky Way halo,
specifically because they have fallen into the Milky Way galaxy only relatively
recently. The general result is that late-falling WIMPs reach the solar
neighborhood as streams of cold dark matter particles with small velocity
dispersion and high bulk velocity. Only few of these streams are expected, each
one coming from a particular direction.

In this paper we discuss the possibility of detecting ``extragalactic WIMPs.''
  These would appear as streams of WIMPs passing the
solar neighborhood.  The existence of a separate dark matter population with
properties distinct from that of galactic WIMPs is plausible although not
certain. Its presence depends on the prehistory and the evolution of the
galaxy. So verification of such an extragalactic component would provide very
interesting information on the history and dynamics of the galaxy.

We address the question of the detectability of extragalactic WIMPs within a
simple model in which the galaxy is moving through a background field of
ambient WIMPs. In this way, there is a simple distinction between bound and
extragalactic WIMPs. Although this is not the commonly used picture at the
present time -- where the galaxy is still in a state of formation and there is
no clear distinction between bound and extragalactic WIMPs -- our infalling
ambient WIMPs may resemble the streams found in detailed simulations. Indeed,
the investigation of our simplified model provides the opportunities to look at
possibilities for detection and to begin to investigate this complex of
questions.

Any such extragalactic WIMPs reaching the Earth from outside the halo will have
been accelerated considerably by falling in the potential well of the Milky
Way.  They may thus be expected to have the local escape velocity from the
Galaxy, roughly 600 km/s, to be compared with the 200-300 km/s typical of halo
WIMPs.  Furthermore, in a picture where the galaxy is moving through a
background of WIMPs essentially at rest, they will appear in the rest frame of
the galaxy as a ``wind'' with definite incoming direction and velocity at
infinity. In the case of a spherically symmetric halo, this leads to their appearance at the Earth from one or two
essentially unique directions, one direction corresponding to WIMPs on their
way into the galaxy, the other to WIMPs on their way out of the galaxy.
Hence in this picture one has extragalactic WIMPs at
the Earth with the distinct features of essentially unique energy and
direction of motion.

In direct detection based on the observation of nuclear recoil, the unique
energy has two interesting implications. First, the recoil energy for
extragalactic WIMPs is higher than that for typical halo WIMPs. Since, in
general, the background decreases sharply with energy \cite{schnee}, this may
be a very favorable point experimentally. Secondly, a monoenergetic incoming
flux produces a characteristic recoil spectrum quite different from the recoil
spectrum from galactic WIMPs. For example, if the WIMP is not too heavy with
respect to the target nucleus, the recoil energy spectrum has a ``box-like''
shape, quite different from the tail of a Maxwellian distribution.

Concerning the definite arrival direction at the Earth, exploitation of this
should eventually be possible with direction-sensitive detectors~\cite{drift}.
The directional sensitivity is of particular interest in connection with
extragalactic WIMPs, and indeed could be used as a filter to increase the
extragalactic WIMP signal relative to halo WIMPs and background.

When directional sensitivity becomes available~\cite{drift}, an improvement of
the signal can be hoped for, in addition to interesting additional information
on the extragalactic WIMPs population.  Unfortunately with present knowledge it
is not possible to calculate this direction at the Earth, since it depends on
the unknown shape of the Galactic gravitational potential (see below).  On the
other hand were a definite direction ever to be verified, this direction, like
the feature of a distinct energy, would provide very interesting information on
the structure and evolution of the galaxy.

Naturally the absolute rate to be expected from extragalactic WIMPs is very
small and this is undoubtedly the major obstacle to a practical realization of
these ideas. However with progress in background reduction and increasing mass
of cryogenic detectors it is perhaps of interest to consider detection of
extragalactic WIMPs for the future in a preliminary way.  We shall find that
under 
the assumption of an extragalactic WIMP population density of the order of the
universal critical density, the expected count rate is very roughly the same as
that of fast halo WIMPs. If there is an enhanced density associated with the
Local Group of galaxies, say by a factor of 10-30 (see below), then the rate is
correspondingly increased. Thus a distinct monoenergetic WIMP population
--if it exists-- could show up as a shoulder above the halo WIMP spectrum.
Verification of such a component would provide very interesting information on
the history and dynamics of the galaxy.

We proceed as follows.  First, in Section~\ref{sec:flux}, we discuss the nature
of the incoming flux of extragalactic WIMPs for an observer in the Solar
System, i.e., the velocity, the direction, and the flux enhancement due to
gravitational focusing.  Then, in Section~\ref{sec:rate}, we discuss
interactions in a detector: cross sections and recoil energy spectrum.  In
Section~\ref{sec:gal}, we discuss the expected density of fast galactic WIMPs,
and how we will use it as a basis for comparison.  In Section~\ref{sec:crit},
we discuss the expected count rates due to our galaxy moving through a critical
density of extragalactic WIMPs.  Finally, in Section~\ref{sec:lg}, we discuss
the expected count rate due to Local Group WIMPs overdense by a factor of 10-30
relative to critical.  We present our numbers relative to the expected count
rates for fast halo WIMPs; the precise count rates depend on the details of the
detectors, details which are the same for Halo and extragalactic WIMPs.

A general note for this paper: given the level of uncertainty tolerated, we
neglect the small velocity of the Earth in its motion around the Sun, and also
neglect the peculiar velocity of the Sun with respect to a frame rotating with
the galaxy at the Sun location (the so-called local standard of rest). In other
words, we take the Earth and the Sun as moving at the rotation velocity of the
galaxy at the Sun's position, $v_{\rm rot} = 220$ km/s. With these
considerations in mind, we will refer to Earth, Sun, and Solar System
interchangeably.  This simplifying assumption precludes us from considering the
annual modulation of the signal due to the Earth motion around the Sun
\cite{dfs,ffg}. We only
comment that the amplitude of this modulation is approximately equal to the
ratio of the Earth's speed, 30 km/s, to the WIMP speed, and so decreases as
the speed of the WIMPs increases. Hence for extragalactic WIMPs arriving on
Earth at the escape velocity, the amplitude of the modulation is expected to be
approximately 5\%.

\section{LOCAL FLUX OF EXTRAGALACTIC WIMPS}
\label{sec:flux}

In this section, we discuss the flux of extragalactic WIMPs as seen by an
observer moving with the Solar System.  The WIMP flux is an essential quantity
for estimating the total interaction rate, which is proportional to the flux
$F$ and the interaction cross section $\sigma$,
\begin{equation}
  R_{tot} = F \sigma ,
\end{equation}
where $F = n w$ is the product of WIMP number density times velocity.
However, the WIMP speed itself is also significant in that it governs the
recoil energy in the detector, and so determines the differential interaction
rate (see Section~\ref{sec:rate}).  Furthermore, given the prospects for a
directional WIMP detector~\cite{drift}, also the direction of the WIMP velocity
is of interest.

{\it Velocity:} When an extragalactic WIMP enters the
Milky Way, it is accelerated to at least the local escape velocity $v_{\rm
  esc}$ by the potential well of our galaxy.  The local escape velocity from
the Milky Way at the position of the Sun is not well determined.  Following the
discussion in \cite{ffg}, we take a value of about $v_{\rm esc} = 600$km/s. The
local speed $v$ of an extragalactic WIMP with velocity $v_\infty$ far away from
the Milky Way follows from energy conservation
\begin{equation}
v = \sqrt{ v_\infty^2 + v_{\rm esc}^2 }   .
\end{equation}
Since we anticipate that WIMPs outside the galaxy have rather low velocities
compared to the escape speed, to a first approximation WIMPs of extragalactic
origin at the Earth will have a unique velocity of $v \sim v_{\rm esc}$ in the
rest frame of the galaxy. In a simple model we consider below, the increase in
$v$ over $v_{\rm esc}$ due to an initial velocity of the WIMPs is on the order
of a few percent.

The transformation from the WIMP velocity in the galactic rest frame to the
WIMP velocity relative to the Earth is obtained by a simple vectorial addition
of its velocity relative to the galaxy and the galactic rotation velocity. As
such, it depends on the arrival direction of the WIMP when it approaches the
Solar System. We define $\psi$ to be the angle between the following two
vectors: the vector starting at the Earth and pointing
in the direction the WIMPs are moving near the Earth,
and the vector starting at the Earth and pointing in
the direction of the galactic rotation velocity. Then
the WIMP speed relative to the Sun is
\begin{equation}
  \label{eq:w}
  w = \sqrt{ v^2 + v_{\rm rot}^2 - 2 v v_{\rm rot} \cos \psi} .
\end{equation}

{\it Arrival Direction:}
Statements on the arrival direction of extragalactic WIMPs are model dependent.
We will focus on a plausible model in which the Milky Way is in motion with
respect to the extragalactic WIMPs in its vicinity.  Viewed in the rest frame of
the Milky Way, there is then a parallel flux of extragalactic WIMPs entering
the galactic halo with a definite velocity at infinity $v_{\infty}$.
(Section~\ref{sec:lg} discusses a particular model of this type in more
detail.)

The further progress of the particles, and in particular their direction upon
reaching the Sun, depends very much on the details of the potential in the
galactic halo. In a flattened halo, WIMPs may follow complicated non-planar
trajectories.  We will consider instead a spherical potential.
In this case, conservation of
energy and angular momentum impose that the WIMP trajectories lie on the
plane defined by the Sun, the direction of galactic motion, and the galactic
center. Spherical symmetry would also impose that only a discrete number of
trajectories on the orbital plane intersect the location of the Earth.  Thus
the detector should see WIMPs coming from a few unique directions, which would
be somewhat blurred by the velocity dispersion of the extragalactic WIMPs.  On
the one hand, it is not easy to predict what these directions are, because they
strongly depend on the radial behavior of the galactic gravitational potential.
(Some examples will be given later in Section~\ref{sec:lg} and in the
Appendix.)  On the other hand, detecting and determining the direction of
possible extragalactic WIMPs would provide very interesting information on the
gravitational potential of the galactic halo.

{\it Focusing: }  A further interesting effect of the galactic
potential is gravitational focusing, which will tend to increase
the flux at
the Sun relative to its value at infinity.
Again, specific results even in the simple model
depend on the details of the galactic potential.

However it is interesting that one may find a simple result for a kind of
average enhancement by the following argument.  We view the galaxy, in its rest
frame, as being subject to a parallel flux of particles, which at great
distances is $F_{\infty}=\rho_{\infty}v_{\infty}/m$, where
$\rho_{\infty} = \rho_{\rm extragal}$ is the density of extragalactic WIMPs far away from the Milky
Way and $m$ is the WIMP mass.  The flux is the appropriate quantity to study
since it is divergenceless in a stationary situation.  Consider a particle
incident at very great impact parameter. In its trajectory through the galaxy,
or rather the galactic halo, it will remain far from the center and suffer a
very small deflection.  As we reduce the impact parameter, the trajectory will
pass closer and closer to the center. Consider the sphere around the center of
the galaxy on which the Sun is located, of radius $R_{\rm Sun}$. As the
incident impact parameter is reduced, we will reach a critical value of the
impact parameter where the trajectory of the particle just touches the sphere
of radius $R_{\rm Sun}$.

Now every incident particle with less than this impact parameter,
which we call $b_{\rm max}$, will pass through the sphere.
Hence the total number of particles
per
second passing through the entire surface of the sphere is
\begin{equation}
N_{\rm sphere}=4\pi b^2_{\rm max}F_{\infty}
\end{equation}

 If we divide this by the surface area of the sphere we will
obtain
the  average flux over the sphere:
\begin {equation}\label{fbar}
\bar F_{\rm sphere}={N_{\rm sphere}\over 4\pi R^2_{\rm Sun}}=
{b^2_{\rm max}\over R^2_{\rm Sun}}F_{\infty}
\end{equation}

We now use angular momentum conservation to find $b_{\rm max}$.
The angular momentum per unit mass as calculated at infinity
for
the particle
on the critical trajectory is $v_{\infty}b_{\rm max}$. On the
other
hand when it just touches the sphere it is $v R_{\rm Sun}$.

We thus obtain
\begin{equation}
b_{\rm max}={v\over v_{\infty}} R_{\rm Sun}
\end{equation}
and the average flux
\begin {equation}\label{fbara}
\bar F_{\rm sphere}=F_{\infty} {v^2\over v^2_{\infty}} .
\end{equation}
This equation is invalid when $v_\infty=0$, in which case there is
no WIMP wind and the simplified model described here should be amended by
including the WIMP velocity dispersion.)

We can also express this result in terms of WIMP densities instead of fluxes.
{}From $\bar{F}_{\rm sphere} =\bar{\rho}_{\rm sphere} v/m$ and $F_{\infty} =
\rho_{\infty} v_{\infty}/m$, it follows that the WIMP density averaged over the
sphere of radius $R_{\rm Sun}$ is given by
\begin{equation}
  \bar{\rho}_{\rm sphere} = {\rho_{\infty}} \frac{v}{v_{\infty}}
\end{equation}
With $v \approx v_{\rm esc}\approx 600$ km/s and $v_{\infty}\approx 75$ km/s
(see Section~\ref{sec:lg}), we obtain a magnification factor $\bar{\rho}_{\rm
  sphere}/ \rho_{\infty}\approx 8$.

This is an essentially exact result within the model but unfortunately it is
only the average over the entire sphere in question, while instead we would
like to know the density of extragalactic WIMPs at the location of the Solar
System $\rho_{\rm extragal,local}$. This density depends on the details of the
gravitational potential, and is in general correlated with the arrival
direction of the extragalactic WIMPs at the Sun.  Obtaining the exact value of
the enhancement factor at the position of the Earth is quite complicated. Given
the greater uncertainties of other numbers here, this calculation is not
currently justified. We therefore parametrize the uncertainty in the focusing
enhancement by a factor $b_{\rm foc}$, and write
\begin{equation}
  \label{eq:focenh}
  \rho_{\rm extragal,local} = b_{\rm foc} \bar{\rho}_{\rm sphere} .
\end{equation}
Since the position of the Sun on this sphere is not exceptional\footnote{As
  discussed in Section~\ref{sec:lg}, our best guess is that WIMPs enter the
  Galaxy from the direction of Andromeda, which lies $\sim 120^\circ$ from the
  Galactic center as seen from the Solar System.}, the average should roughly
characterize the enhancement factor to be expected.  In more detailed
calculations using various potential shapes, we find there can be variations of
two or three in the enhancement factor but, barring very special conditions,
not of an order of magnitude; i.e., we find $ 0.3 \lesssim b_{\rm foc} \lesssim
3 $.

\section{INTERACTION RATE IN A DETECTOR}
\label{sec:rate}

The total interaction rate of WIMPs with a target detector nucleus is given by
\begin{equation}
\label{eq:totrate}
R_{\rm tot} = F \sigma ,
\end{equation}
where $F=\rho w/m$ is the WIMP flux, $m$ is the WIMP mass,
and $\sigma$ is the cross section for the
interaction.  Since the flux of Milky Way WIMPs largely dominates over any
imaginable flux of extragalactic WIMPs, the total interaction rate in a
detector is expected to be largely dominated by galactic, instead of
extragalactic, WIMPs.

However, as seen in Section~\ref{sec:flux}, the local speed of an
extragalactic WIMP, and so its kinetic energy, is higher than that of any
galactic WIMP. In scattering with a detector nucleus, the more energetic
extragalactic WIMP can deposit more energy, and so impart a higher recoil
energy to the detector nucleus than any galactic WIMP. Hence selecting events
with high recoil energy helps in searching for a signal due to extragalactic
WIMPs. It is therefore important to consider the interaction rate differential
in the nucleus recoil energy, and compare this differential rate for
extragalactic WIMPs with that expected from fast galactic WIMPs. In the rest of
the Section, we review the generalities of the differential interaction rate,
that is the recoil spectrum, and in the following sections we discuss and
compare the recoil spectra expected from extragalactic and fast galactic WIMPs.

In general, the differential cross section for elastic scattering of a WIMP of
mass $m$ with a nucleus of mass $M$ can be written
\begin{equation}
\label{eq:diffcross}
{d\sigma \over dq^2} = {\sigma_0 \over 4 \mu^2 w^2} {\cal F}(q)
\end{equation}
where $w$ is the speed of the WIMP relative to the detector, $q$ is the
momentum transfer in the elastic collision, $\mu = mM/(m+M)$ is the
WIMP--nucleus reduced mass, and ${\cal F}(q)$ is a nuclear form factor that
takes into account the loss of coherence in WIMP-nucleus interactions for
momentum transfers comparable to or larger than the inverse nuclear radius (we
normalized ${\cal F}(0) = 1$).

For purely scalar interactions,
\begin{equation}
\sigma_{0,\rm scalar} = {4 \mu^2 \over \pi} [Zf_p + (A-Z)f_n]^2
\,
.
\end{equation}
Here $Z$ is the number of protons, $A-Z$ is the number of
neutrons,
and $f_p$ and $f_n$ are the neutralino couplings to nucleons.
For purely spin-dependent interactions,
\begin{equation}
\sigma_{0,\rm spin} = (32/\pi) G_F^2 \mu^2 \Lambda^2 J(J+1) \, .
\end{equation}
Here $J$ is the total angular momentum of the nucleus and $\Lambda$ is
determined by the expectation value of the spin content of the nucleus (see
\cite{goodman,dfs,epv,ressell,jkg}).

The nucleus recoil energy equals the energy lost by the WIMP and is simply
\begin{equation}
\label{eq:recoil}
E = \frac{q^2}{2M} .
\end{equation}
The maximum momentum transfer is $q_{\rm max} = 2 \mu v$, and the maximum
recoil energy imparted by a WIMP of speed $w$ follows as
\begin{equation}
  \label{eq:emax}
  E_{\rm max} = \frac{ 2 \mu^2 w^2 } { M} .
\end{equation}

The recoil spectrum for WIMPs of a given speed $w$ then follows from a change
of variables as
\begin{equation}
  \label{eq:ratesingle}
  \frac{dR}{dE} = 2 F {d\sigma \over dq^2} =
  \left\{ \begin{array}{ll}
      \displaystyle \frac{\rho \sigma_0}{2 \mu^2 m w} {\cal F}(q) ,
      & {\rm for~} E \le E_{\rm max},
      \\ 0 ,
      & {\rm for~} E > E_{\rm max} .
  \end{array} \right.
\end{equation}
We have used $F=\rho w/m$, and we have divided by the nucleus mass so as to
obtain a rate per unit detector mass, as customary.

As argued in Section~\ref{sec:flux}, extragalactic WIMPs in the solar
neighborhood have essentially a single speed $w$. Inserting the focusing
enhancement factor as in eq.~(\ref{eq:focenh}), we obtain that extragalactic
WIMPs give rise to the following recoil spectrum
\begin{equation}
  \label{eq:drde-extragal}
  \left. \frac{dR}{dE} \right|_{\rm extragal} =
  \left\{ \begin{array}{ll}
      \displaystyle
      \frac{b_{\rm foc} }{w } \,
      \frac{v }{v_{\infty} } \,
      \frac{\rho_{\infty} \sigma_0}{2 \mu^2 m} \,
      {\cal F}(q) ,
      & {\rm for~} E \le E_{\rm max},
      \\ 0 ,
      & {\rm for~} E > E_{\rm max} .
  \end{array} \right.
\end{equation}
Here $w$ is given by the expression in eq.~(\ref{eq:w}). Notice that both the
focusing enhancement factor $b_{\rm foc}$ and the WIMP speed relative to the
Sun $w$ depend on the details of the galactic gravitational potential.

The nuclear form factor ${\cal F}(q)$ depends on the type of WIMP-nucleus
interaction and on the mass and spin distributions within the nucleus. Some of
these distributions can be found in \cite{epv,ressell,jkg}.  For the present
simple considerations, we consider only the case where form factor effects are
negligible, $q \ll \hbar / R$, where $R$ is the radius of the nucleus. (For a
velocity of 600 km/s this translates into either $m \ll M/[(Mc^2/20{\rm
  GeV})^{4/3} -1]$ or $M\lesssim 20$ GeV/$c^2$ independently of $m$.)  The
cross section is then simply $\sigma_0$.

With extragalactic WIMPs all moving at a single velocity and form factor
effects neglected, the differential cross section is constant up to the
kinematic maximum. Thus from eqs.~(\ref{eq:diffcross}) and (\ref{eq:recoil}),
for low mass extragalactic WIMPs, the recoil spectrum in a detector is
``box-like'', i.e.\ $dR/dE=(\rho \sigma_0)/(2 \mu^2 m w) = {\rm constant}$ is
flat as a function of recoil energy $E$ up to the maximum recoil energy $E_{\rm
  max}$ given in eq.~(\ref{eq:emax}).  This signature is completely different
from that obtained from a Maxwellian velocity distribution; for halo WIMPs, one
must average these boxes over different velocities. We recall the procedure in
the next Section.

\section{COMPARISON WITH FAST  GALACTIC WIMPS}
\label{sec:gal}

It is useful to have a simple comparison of the expected detection rates for
extragalactic and galactic WIMPs. For galactic WIMPs, we use the detection
rates previously calculated by many authors as a basis for the comparison. In
more detail, we compare the extragalactic rates with the galactic rates
obtained under the simple assumption that the galactic WIMP speeds obey a
Maxwellian distribution in the rest frame of the galaxy, without truncating it
at the escape velocity. This is the WIMP velocity distribution used in the
earliest analyses of direct detection by Drukier, Freese, and Spergel
\cite{dfs} and Freese, Frieman, and Gould \cite{ffg}. We write it as
\begin{equation}
  \label{eq:max}
  f^{\rm gal}_{\rm MW}(v)dv = 4 \pi \rho_{\rm MW}
  \left[{3 \over 2 \pi \bar{v}^2 }\right]^{3/2}
  v^2 \, \exp\left[-{3 v^2 \over 2 \bar{v}^2} \right] dv ,
\end{equation}
where we take the dispersion velocity $\bar{v} = 270$ km/s.  Assuming an
isothermal sphere, the local galactic WIMPs have a density of $\sim 0.3$
GeV/cm$^{3}/c^2$, which amounts to
\begin{equation}
  \label{eq:MWdens}
  \rho_{\rm MW} \sim 60,\!000 ~\rho_{\rm crit}
\end{equation}
(for a Hubble constant of 70 km s$^{-1}$ Mpc$^{-1}$).  Here $\rho_{\rm crit} =
5 \times 10^{-6} h_{70}^2 {\rm GeV}/{\rm cm^3}/c^2 $ where $h_{70}$ is the
Hubble constant in units of 70 km/s/Mpc.

The distribution in eq.~(\ref{eq:max}) that we use as a basis for comparison,
although useful for this purpose, may not be the real distribution of WIMP
velocities, first, because it allows galactic WIMPs with speeds in excess of
the escape speed, and second, because the real nature of the WIMP velocity
distribution in the galaxy is unknown.  The effects of other halo distributions
on direct detection have been studied for example in
refs.~\cite{copi,green,gelminigondolo,stiff,kamion}.

We now present two ways to estimate the ratio of extragalactic to galactic
event rates. One method gives an order of magnitude estimate, the other is more
detailed, but they give essentially the same result.

The quick method is based on the comparison of the galactic and extragalactic
WIMP fluxes at the escape speed in the galactic rest frame, $v_{\rm esc}
\approx 600$ km/s. How many galactic WIMPs are moving this fast?  We use the
Maxwellian velocity distribution of eq.~(\ref{eq:max}) to find that the
fraction of fast galactic WIMPs (out of the total galactic WIMPs) that are
within $\Delta v=1$ km/s of escape velocity (corresponding to 1\% energy
resolution) is given by
\begin{equation}
  \label{eq:600frac}
  \rho_{\rm MW,esc} \simeq
  f^{\rm gal}_{\rm MW}(v_{\rm esc}) \Delta v \sim 4
  \times 10^{-5} \rho_{\rm MW}
\end{equation}
so that
\begin{equation}
  \label{eq:rho600}
  \rho_{\rm MW, esc} \sim 3 \, \rho_{\rm crit} .
\end{equation}
With a reasonable guess of 3 for the focusing enhancement
factor, the local density of extragalactic WIMPs is
\begin{equation}
  \rho_{\rm extragal, local} \sim 3 \, \rho_{\infty} .
\end{equation}
Hence
\begin{equation}
  \rho_{\rm extragal, local} \sim \rho_{\rm MW, esc} \,
  \frac{\rho_{\infty}}{\rho_{\rm crit}} .
\end{equation}
The ratio of interaction rates for extragalactic and galactic WIMPs moving with
escape speed then follows as
\begin{equation}
  \label{eq:quickratio}
  \frac{ R_{\rm extragal} } { R_{\rm MW, esc}} \sim
  \frac{\rho_{\infty}}{\rho_{\rm crit}} .
\end{equation}
We remind the reader that, in the above equation, we have used the integrated
rates for extragalactic and galactic WIMPs, with the proviso that for the
galactic WIMPs the rate is only integrated over velocities near local escape
speed (and hence is {\it not} the total rate integrated over all energies).

We now discuss the more detailed method of comparing the galactic and
extragalactic rates. Here we calculate the ratio of differential rates rather
than of the integrated rates as given in eq.~\ref{eq:quickratio}.
First, we convert the speed distribution in the galactic
rest frame to a speed distribution as seen by an observer moving with the Solar
System. We do this by means of a Galilean transformation of the velocities.
Denoting with $w$ the WIMP speed relative to the Sun, and remembering that we
identify the velocity of the Sun with the velocity of galactic rotation $v_{\rm
  rot}$, we find
\begin{equation}
  f_{\rm MW}(w) dw =
  2 \rho_{\rm MW}
  \left[\frac{3}{2 \pi \bar{v}^2}\right]^{1/2}
  \frac{w}{v_{\rm rot}} \,
  \sinh\!\left[\frac{3 w v_{\rm rot}}{\bar{v}^2}\right]
  \, \exp\left[-\frac{3(w^2+v_{\rm rot}^2)}{2\bar{v}^2}\right]
  dw .
\end{equation}

The differential interaction rate is then obtained by integrating
eq.~(\ref{eq:ratesingle}) over the WIMP speed distribution $f(w)$. In general,
\begin{equation}
\label{eq:diffrate}
  \frac{dR}{dE} = {\rho \sigma_0 \over 2 \mu^2 m} {\cal F}(q)
  \int_{w_E}^\infty \frac{f(w)}{w} dw ,
\end{equation}
where $w_E = \left( E M/2 \mu^2 \right)^{1/2}$ is the minimum velocity of the
WIMP required to deposit an amount of energy $E$.  For the Maxwellian we use as
comparison, we find
\begin{equation}
  \label{eq:drde-gal}
  \left. \frac{dR}{dE} \right|_{\rm MW} =
  \frac{\rho_{\rm MW} \sigma_0}{2 \mu^2 m} \, {\cal F}(q) \,
  \frac{1}{2 v_{\rm rot}} \left\{
    \erf\left[ \frac{ \sqrt{3} (w_{E} +
        v_{\rm rot})}{\sqrt{2}\bar{v}}\right]
    -\erf\left[ \frac{ \sqrt{3} (w_{E} -
        v_{\rm rot})}{\sqrt{2}\bar{v}}\right]
      \right\} .
\end{equation}

The ratio between the extragalactic and galactic recoil spectra $dR/dE|_{\rm
  extragal}$ and $dR/dE|_{\rm MW}$ depends on the recoil energy and on the
arrival direction of the extragalactic WIMPs. For our comparison, we adopt
the maximum value of this ratio of differential rates,
\begin{equation}
  {\cal R}^{\rm max} = \max \frac{ dR/dE|_{\rm extragal} }{ dR/dE|_{\rm MW} } .
\end{equation}

Clearly the ratio of differential rates is maximal when the recoil energy
equals the maximum recoil energy $E_{\rm max}$. Furthermore, if we neglect the
dependence of the focusing factor on the arrival direction, the local
extragalactic flux is maximal when the WIMP velocity is opposite to the
direction of galactic rotation, that is $\psi=\pi$ in the notation of
Section~\ref{sec:flux}. This is therefore the value of ${\cal R}^{\rm max}$
that we adopt for our comparison. Explicitly, we use
\begin{equation}
  \label{eq:rmax}
  {\cal R}^{\rm max} =
  b_{\rm foc} \,
  \frac{v}{v_{\infty}} \,
  \frac{\rho_{\infty} } {\rho_{\rm MW} } \,
  \frac{2 v_{\rm rot}}{w_{\rm max}} \,
  \frac{ 1 }
  {\displaystyle \erf\left[ \frac{ \sqrt{3} (w_{\rm max} +
        v_{\rm rot})}{\sqrt{2}\bar{v}}\right]
    -\erf\left[ \frac{ \sqrt{3} (w_{\rm max} -
        v_{\rm rot})}{\sqrt{2}\bar{v}}\right] }
\end{equation}
where $ w_{\rm max} = v_{\rm rot} + v $ and $v=\sqrt{ v_{\infty}^2 + v_{\rm
    esc}^2 } $.  For $v_{\infty} \approx 75 $ km/s (see Section~\ref{sec:lg}),
and with the values adopted for escape speed, rotation velocity, and velocity
dispersion, the ${\cal R}^{\rm max}$ ratio becomes
\begin{equation}
 {\cal R}^{\rm max} \sim 40,\!000 \, b_{\rm foc} \,
 \frac{\rho_{\infty} } {\rho_{\rm MW} },
\end{equation}
and using eq.~(\ref{eq:MWdens}),
\begin{equation}
  \label{eq:maxratio}
  {\cal R}^{\rm max} \sim 0.7 \, b_{\rm foc} \,
  \frac{\rho_{\infty} } {\rho_{\rm crit} }.
\end{equation}
This is the ratio that will appear in Figure 1 below.
This ratio of differential rates
 is within a small factor equal to the analogous expression for
the ratio of integrated rates in eq.~(\ref{eq:quickratio})
(integrated only over velocities near the escape speed).

We conclude that within an order of magnitude the ratio of extragalactic to
galactic WIMP-detection rates is
\begin{equation}
  \label{eq:ratio}
  \frac{\rm rate(extragalactic)}{{\rm rate(galactic~near~}v_{\rm esc}{\rm )}}
  \sim
  \frac{\rho_{\infty}}{\rho_{\rm crit}} .
\end{equation}

A few comments are in order.  The detection rate of galactic WIMPs used for
comparison in eqs.~(\ref{eq:quickratio}), (\ref{eq:maxratio}) and
(\ref{eq:ratio}) is a fictitious extrapolation of the detection rate of
galactic WIMPs to WIMP velocities exceeding the escape speed. By definition, no
galactic WIMPs, i.e.\ no WIMPs bound to the galaxy, exist with speeds above the
escape speed. The comparison ratios in eqs.~(\ref{eq:quickratio}),
(\ref{eq:maxratio}) and (\ref{eq:ratio}) should then be understood not as
actual ratios of detection rates at specific recoil energies, but only as an
informative way to gauge the rates due to extragalactic WIMPs with respect to
the galactic signal.  Indeed, the ratio of extragalactic to galactic
WIMP-detection rates may in reality be much larger than what
eq.~(\ref{eq:ratio}) indicates: The galactic WIMPs are limited to speeds
smaller than the escape velocity, while the extragalactic WIMPs may have speeds
higher than the escape velocity. Hence the upper edge of the ``box-like''
spectrum of extragalactic WIMPs may extend beyond the highest recoil energy
from galactic WIMPs. This may make the extragalactic signal stick out above the
galactic signal more than the numbers in this section (and below) would
indicate. On the other hand, the ratio of extragalactic to galactic rates may
be much smaller than what eq.~(\ref{eq:ratio}) indicates: The shape of the
galactic potential or the motion of the galaxy with respect to the
extragalactic WIMPs may conspire to make the extragalactic WIMPs move in a
direction almost parallel to the direction of galactic rotation at the location
of the Sun in the galaxy. In this case, the speed of extragalactic WIMPs
relative to the Sun would be a low 380 km/s, and the recoil spectrum of
extragalactic WIMPs would be much lower than the recoil spectrum of galactic
WIMPs even at the highest recoil energy. This is an unfortunate possibility
than should be kept in mind. Hopefully, a detector with directional sensitivity
may be able to help in this situation.

\section{EXTRAGALACTIC WIMPS WITH CRITICAL DENSITY}
\label{sec:crit}

We consider smoothly distributed extragalactic WIMPs with density equal to the
critical density $\rho_{\rm crit}$. We expect that the typical velocities of
these WIMPs away from our Galaxy  are small compared to the local
escape velocity from the Milky Way.  Applying the considerations in the
previous sections, we find that the density of extragalactic WIMPs near the
Earth is comparable to the density of galactic WIMPs moving at close to escape
velocity (600 km/s),
\begin{equation}
\label{eq:rationhw}
\rho_{\rm crit,local} \sim \rho_{\rm MW,600} .
\end{equation}
The subscript (crit,local) is intended to remind the reader that this is the
local density of extragalactic 
WIMPs that at infinity have a critical density.  We find a
similar relation for the detection rates,
\begin{equation}
R_{\rm crit} \sim R_{\rm MW,600};
\end{equation}
that is, the expected extragalactic WIMP count rate is comparable to that for
fast halo WIMPs moving at close to escape velocity.

However, these estimates are very uncertain, because, not knowing the relative
motion of the galaxy with respect to critical-density extragalactic WIMPs, the
considerations at the end of the Section~\ref{sec:gal} unfortunately apply
fully. A somewhat less uncertain case is discussed next.

\section{LOCAL GROUP WIMPS}
\label{sec:lg}

It would be interesting if there would be an overdensity (above critical) of
extragalactic WIMPs near the Milky Way. If so, these extragalactic WIMPs would
have a count rate in excess of the count rate from halo WIMPs at escape
velocity.  For example, our galaxy may be sitting in an overdensity of WIMPs
due to the fact that we are in the Local Group of galaxies.  If the Local Group
would be overdense relative to critical by a factor of 10-30, then the expected
count rate in a detector would be 10-30 times that of fast halo WIMPs near the
escape velocity.  In addition, if the Milky Way is moving in the direction of
Andromeda (as argued below), then we would be moving into a wind of Local Group
WIMPs. In this case, the Local Group WIMPs would
approach our detectors from one or at most two unique directions. Here we
consider this possibility.

The Local Group of galaxies is a group of galaxies dominated by the Milky Way
and M31 (a.k.a.\ the Andromeda galaxy).  Currently the number of galaxies
stands at 35.  For a discussion of Local Group properties, see
\cite{vandenbergh,mateo}.  Each of these big galaxies has a few small
satellites (these include the Large and Small Magellanic Clouds for the Milky
Way).  In addition, there are some isolated small galaxies.  Most of the mass
of the Local Group is concentrated in the Milky Way and Andromeda. The Milky
Way is roughly 400 kpc from the center of gravity of the Local Group.  The zero-velocity
surface, which separates the Local Group from the field that is expanding with
the Hubble flow, has a radius 1.2-1.8 Mpc from the center of gravity of the
Local Group.

The question is, are there any Local Group WIMPs that are not virialized with
the Milky Way?  We will stipulate the existence of a $10^{12} M_{\odot}$ Local
Group WIMP halo (not virialized), with the caveat that in the worst case there
may be no such halo at all.  Recent studies of galaxies
\cite{bernstein,kaiser,sloan} indicate that galaxies are larger than previously
believed, with radii at least 250-300 kpc, and mass-to-light ratios of 145-200.
Hence it is possible that much of the dark matter in the universe is
concentrated in the galaxies themselves.  However, even if this is the case, it
is plausible to believe that when galaxies come into contact with one another
some of the dark matter is stripped off.  Hence it is possible that there are
Local Group WIMPs not associated with either the Milky Way or M31; these form a
Local Group halo but are not virialized.  In particular, explaining the rapid
motions of galaxies near the outskirts of the Local Group may require the
existence of a Local Group halo not associated with either the Milky Way or
Andromeda galaxy.  Hence in the remainder of this section we will stipulate the
existence of a $10^{12} M_{\odot}$ Local Group halo of WIMPs.

We now specialize the considerations of the previous Sections to Local Group
WIMPs.

{\it WIMP velocity and arrival direction:} The velocity dispersion within the
Local Group is $\sigma_r = 61 \pm 8$ km s$^{-1}$ \cite{vandenbergh}.  The Milky
Way is moving towards the center of gravity of the Local Group with a similar
velocity (about 75 km/s).  The tangential velocity of the Milky Way towards the
Local Group has not been measured.  However, it is likely to be small: the
Local Group is an overdense region that was expanding and now collapsing, so
that the Milky Way is likely to be falling into M31.  Hence we can assume the
Milky Way is moving in the direction of M31 with a speed of 75 km/s. This gives
rise to a WIMP wind from the direction of M31 with a velocity at infinity
$v_{\infty} \approx 75 $ km/s.

The Local Group dispersion velocity of 60 km/s is only 10 percent of the Milky
Way escape velocity of 600 km/s, and the associated energy is only 1\%.  Hence
energy conservation tells us we need to modify the velocity near Earth by only
1\%.  Thus as above, we find that, with respect to the galactic rest frame, the
speed of Local Group WIMPs in the solar neighborhood is to a good approximation
$\sim 600$ km/s, that is the escape speed from the galaxy.

As discussed in Section~\ref{sec:flux}, the specific arrival direction of
extragalactic WIMPs at the location of the Sun and Earth depends on the details
of the galactic gravitational potential. For the sake of illustration, we have
assumed a spherical Milky Way halo and we have calculated WIMP trajectories for
two radial dependences of the gravitational potential: a Navarro-Frenk-White
potential and a logarithmic potential. (See the appendix for details.) We have
found that at most two arrival directions at the Sun are possible, one for
WIMPs on their way into the Milky Way and one for WIMPs on their way out. Thus
there are one or at most two values of the WIMP speed at Earth.

Without any constraint on the direction and speed of the relative motion of the
Milky Way and the Local Group WIMPs, these directions could cover the whole
possible range from opposite to parallel to the Sun motion in the galaxy (which
is roughly the direction of galactic rotation). In this case, the WIMP speed
relative to the Earth, obtained subtracting the Sun velocity from the WIMP
velocity vectorially, could have a value (or at most two values) in the range
380--820 km/s, which is 600$\pm$220 km/s, according to the arrival direction at
the Sun.

However, if we assume that the WIMPs are approaching the galaxy at infinity
from the direction of M31, we find that the WIMP speed relative to the Earth
must be larger than $\sim 640$ km/s. This gives a higher ratio of Local Group
and galactic WIMPs counting rates. The reason for the selection of the higher
range in speeds is the conservation of the sign of angular momentum. To keep
track of the angular momentum, it is helpful to introduce an observer at the
Earth that faces the galactic center with the north galactic pole above his/her
head. Then M31 lies in the southern hemisphere on his/her left-hand side, at
$\sim 120^\circ$ from the galactic center (M31 has galactic longitude
$l=121\fdg2$ and galactic latitude $b=-21\fdg6$). Now consider the WIMPs that
come from M31 and that reach the Earth on their way {\it into} the galaxy, that
is {\it before} they reach the point of closest approach to the galactic
center.  Their initial angular momentum points toward the north galactic pole.
When these WIMPs arrive at the solar neighborhood, their angular momentum still
points toward the north galactic pole (conservation of the sign of angular
momentum).  So for our hypothetical observer they possess a velocity component
(tangential velocity) directed toward the {\it right}, together with another
velocity component (radial velocity) directed toward the galactic center. The
key point is that the direction of galactic rotation, and also the direction of
the solar motion, is to the {\it left} of the observer. Hence the angle between
the WIMP velocity (sum of radial and tangential components) at the solar
neighborhood and the direction of galactic rotation cannot be smaller than
$90^\circ$. It follows that the WIMP speed at the Earth cannot be smaller than
$(600^2+220^2)^{1/2} = 640$ km/s. A more detailed analysis of the geometry also
shows that the angle between the WIMP velocity at the Earth and the direction
of galactic rotation cannot be larger than $155^\circ$, which gives a maximum
WIMP speed relative to the Earth of 805 km/s.

{\it Density at the Earth:} Using the numbers given above for the total mass
$10^{12} M_{\odot}$ in WIMPs and a Local Group radius of 1.2-1.8 Mpc, we find
an average Local Group WIMP mass density of
\begin{equation}
\label{eq:rhodm}
\rho_{\rm LGW} = (1-3) \times 10^{-28} \, {\rm gm}/{\rm cm}^3
\,
.
\end{equation}
For a Hubble constant of 70 km s$^{-1}$ Mpc$^{-1}$, this amounts to
(10-30)$\rho_{\rm crit}$.  Our galaxy is roughly 400 kpc from the center of
gravity of the Local Group, about 1/3 of the way out to the zero-velocity
surface; for a $1/r^2$ density profile, we should thus be at roughly
10$\rho_{\rm crit}$.  (Since the system is not virialized we cannot be sure of
the density profile, but this estimate still gives us an indication that is not
unreasonable to use an average WIMP density in the halo as its value at the
outskirts of the Milky Way.)

Thus the density ratio of nearby Local Group WIMPs to galactic
WIMPs
moving with
600 km/s is given by
\begin{equation}
\label{eq:numberratio}
\rho_{\rm LGW,local}  \sim (10-30) \rho_{\rm MW,600} \, ,
\end{equation}
where $\rho_{\rm LGW,local}$ is the density of Local Group WIMPs near
the
Earth and
$\rho_{\rm MW,600}$ is the density of Milky Way WIMPs moving with 600
km/s.

The count rate of Local Group WIMPs in detectors would also be 10-30 times that
of galactic WIMPs
\begin{equation}
R_{\rm LGW} \sim (10-30)  R_{\rm MW,600}.
\end{equation}
These Local Group WIMPs would have a unique signature in that (i) they come
only from one or at most two specific directions and (ii) the lighter ones
would have a flat recoil spectrum.

Unfortunately the numbers of events in current detectors would still be tiny.
The expected count rate is $0.1$ below that for typical halo WIMPs moving $\sim
220$ km/s, and $10^{-3}$ below the signal from the sum total of halo WIMPs.

However, as discussed in the Introduction and at the beginning of
Section~\ref{sec:flux}, selecting events with high recoil energy should help in
the detection of Local Group WIMPs, because the recoil spectrum of
extragalactic WIMPs would show up as a high-energy shoulder which may stick out
above the spectrum of galactic WIMPs at high recoil energies. We calculate the
recoil spectra for extragalactic WIMPs using eq.~(\ref{eq:drde-extragal}) and
compare it with the galactic spectrum from eq.~(\ref{eq:drde-gal}). As
discussed in Section~\ref{sec:rate}, the spectrum of extragalactic WIMPs is
boxlike.

Figures 1 and 2 show examples of the expected recoil spectra for Milky
Way and Local Group WIMPs. For the sake of illustration, we have fixed the mass
of the target nucleus $M=73$ GeV/$c^2$, the WIMP mass $m=60$ GeV/$c^2$, the
WIMP-nucleus scattering cross section $\sigma_0=10^{-35}$ cm$^2$, and we have
set the nuclear form factor ${\cal F}(q)=1$. For the galactic WIMPs, we have
obtained the recoil spectrum from eq.~(\ref{eq:diffrate}) for a Maxwellian
velocity distribution as in eq.~(\ref{eq:max}) but truncated at the escape
velocity (we do not display the lengthy formulas). For comparison, the dotted
line shows the spectrum due to Milky Way WIMPs for the non-truncated velocity
distribution we use as a comparison in Section~\ref{sec:gal}, that is
eq.~(\ref{eq:drde-gal}).  For the Local Group WIMPs we have obtained the rate
using eq.~(\ref{eq:drde-extragal}), taking $b_{\rm foc}=1$, $\rho_{\rm LGW} =
20 \rho_{\rm crit}$.  Since we do not know the arrival direction of the Local
Group WIMPs at the Solar System, we show two possible spectra for these WIMPs,
Figure~1 and Figure~2. We believe that these two spectra bracket the
plausible range. They correspond to the two values $90^\circ$ and $155^\circ$
for the angle $\psi$ between the velocity of the Solar System in the galaxy and
the velocity of the Local Group WIMPs when they reach the Solar System
(both velocity vectors starting at the earth).

In the most favorable case of $\psi=155^\circ$, the high energy recoils
due to Local Group WIMPs can be clearly distinguished from the recoils due to
galactic WIMPs. In the less favorable case of $\psi=90^\circ$ the
distinction is more difficult.

\section{CONCLUSIONS}
\label{sec:conclusions}

Extragalactic WIMPs have several unique signatures which may aid in their
ultimate detectability.  As these WIMPs enter the Milky Way, they are sped by
the gravitational potential of the Galaxy.  Near the solar neighborhood, they
are moving with the local escape velocity of 600 km/s relative to the galactic
rest frame.  The relatively high velocity of the extragalactic WIMPs can have
the advantage of moving the recoil signal in direct detection to higher
energies, where background is much lower.  In addition, the extragalactic WIMPs
come from only one or two directions.  Although it is currently not easy to
predict what these directions are, because they strongly depend on the radial
behavior of the Galactic gravitational potential, still the uni- or
bi-directionality will someday be a strong signature for an extragalactic WIMP
population.  Additionally, the essentially monochromatic spectrum of
extragalactic WIMPs implies (if they are not too heavy) a nuclear recoil
spectrum that is flat up to the maximum recoil energy.

For a critical density of extragalactic WIMPs, the count rate in
a
detector is roughly the same as that of fast galactic WIMPs, but
their
signature is quite different.  An extragalactic population
of particular interest would be WIMPs associated with the Local
Group;
these may have a 10-30 times higher count rate than galactic
WIMPs
of the same speed, for an energy resolution of 1\% in the
detector.
Detection of extragalactic WIMPs would bring valuable
new insight into the processes of galaxy formation.

We also wish to point out that the annual modulation of the
signal
due to the
Earth moving around the Sun decreases as one goes to higher
velocity WIMPs.
The Earth's annual motion of 30 km/s is 5\% of the escape
velocity
and hence
5\% of the velocity of extragalactic WIMPs.

\section*{Acknowledgments}

We would like to thank Tom Abel, Craig Copi, Catherine Cress, Matt Lewis, Mario
Mateo, Richard Schnee, Lawrence Widrow, and Saleem Zaroubi for very helpful
discussions. We also thank Simon White for discussions on structure formation.
We thank the Department of Energy for support through the Physics Department at
the University of Michigan (contract number DE-FG02-95ER40899) and the
Department of Physics at Case Western Reserve University (contract number
DE-FG02-95ER40898), and the Max Planck Institut f\"ur Physik for support during
the course of this work.

\newpage

\appendix

\section{Trajectories of extragalactic WIMPs.}

In this appendix, we mention some details of the calculation of
the trajectories of extragalactic WIMPs entering the galaxy.
In particular, if a WIMP at infinite distance from the galaxy is
approaching with one angle $\alpha_\infty$, we calculate the angle
$\alpha$ at which it
arrives at the Sun.  The deviation of the angle from its initial
value can be huge.  One might assume that radial infall through
the location of the Sun to the
center of the galaxy is a good approximation to the motion;
however, this assumption is not always true, depending on the
galactic potential.  In an extreme case, the incoming particle
can instead have a purely tangential velocity at the location of the Sun,
with no radial component (towards the Galactic Center) at all.
We considered only galactic potentials which are
spherically symmetric.  Specifically, we present our results for (i) a
Keplerian (point mass) potential, (ii) a
Navarro-Frenk-White (NFW) potential, and (iii) a logarithmic potential.
For comparison, we also consider (iv) the free particle case.

{\it Gravitational potentials:}

(i) As a simple check, we consider a Keplerian potential with
$\phi_{\rm Kep}(r) = - GM/r $, a point mass at the center of the galaxy.
This is {\it not} a realistic potential.

(ii) An NFW potential corresponds to an
NFW density profile $\rho = \rho_s
(r/r_s)^{-1}(1+r/r_s)^{-2}$, where $r_s$ is a length scale and $\rho_s$ is
four times the density at $r=r_s$. Explicitly,
\begin{equation}
\phi_{\rm  NFW}(r) = - 4 \pi G \rho_s r_s^2 \frac{ \ln(1+r/r_s)}{r/r_s} ,
\end{equation}
We fix the scale $r_s$ by imposing that the escape velocity from the galaxy at
the Sun position $r_0$ is $v_{\rm esc}$, a given quantity. This gives the
following equation for $u_s = r_0/r_s$,
\begin{equation}
\label{eq:u_s}
\frac{u_s}{1+u_s} \ln(1+u_s) = 1 - \frac{2 v_c^2}{v_{\rm esc}^2} .
\end{equation}
Here $v_c$ is the galactic circular velocity at the Sun position.  Taking
$v_c=220$ km/s and $v_{\rm esc}=600$ km/s gives $u_s = 2$, in other words
$r_s = r_0/u_s = r_0/2.00$ or $r_s = 4$ kpc if $r_0=8.5$ kpc.  [We are aware
that this value of $r_s$ is an approximation to what may arrive in a realistic
scenario.]

(iii) A logarithmic potential is interesting
in this context because it is a simple
function with constant circular velocity. However, it increases without limit
as $r\to\infty$, and so it needs to be cut off at a certain radius $R$ to allow
for unbound orbits. Hence we use the truncated logarithmic potential
\begin{equation}
\phi_{\rm Log}(r) = \cases{
  \displaystyle v_c^2 \left[ \ln\left(\frac{r}{R}\right) - 1 \right] ,
    & for $r\le R$ , \cr
  \displaystyle - v_c^2 \frac{R}{r} , & for $ r \ge R $ , }
\end{equation}
fixing the cut-off radius $R$ by again imposing that the escape speed at the
Sun position is $v_{\rm esc}$. This gives the condition
\begin{equation}
\label{eq:u_R}
u_R = \frac{r_0}{R} = \exp[1-v_{\rm esc}^2/(2v_c^2)] .
\end{equation}
With $v_{\rm esc}$ and $v_c$ as above, $u_R = 0.0659$.

{\it Arrival directions:} Since we assume spherical symmetry for the galactic
potential, angular momentum is conserved and the orbit is planar. The
particle trajectories that pass by the Sun lie on a single plane, which is
defined by the particle velocity vector at infinity (the stream velocity) ${\bf
  v}_\infty$ and the vector from the galactic center to the Sun ${\bf r}_0$.
[The only exception is if the WIMPs come towards the Sun
from directly behind the galactic center, as discussed below.]

In the plane of the orbit, we define the angles $\alpha_\infty$ and $\alpha$.
To characterize the direction from which the particle entered the Galaxy,
we define the angle $\alpha_\infty$ to be the angle between the
following two vectors that originate at the Galactic Center:
the particle velocity vector at infinity towards the Galactic Center,
and a vector pointing from the Sun towards the Galactic Center
(but translated to the location of the Galactic Center).
To characterize the direction the particle is moving
once it reaches the Sun, we define $\alpha$ as the angle between
the following two vectors that originate at the Sun:
the tangent vector of the particle's motion
as it passes the Sun, and the vector from the Sun towards the Galactic Center.
As an example,
see Figure 4.
In order to obtain $\alpha_\infty$ as a function of $\alpha$, we need
to solve the orbit equations.

Introduce polar coordinates $(r,\theta)$ in the plane of the orbit with the
origin at the Galactic Center.  Let $r_0$ be the distance of the Sun from the
Galactic Center, and $v_\infty$ be the particle velocity at infinity. Then
conservation of energy gives
\begin{equation}
\label{eq:energy}
\frac{1}{2} {\dot r}^2 + \frac{1}{2} r^2 {\dot \theta}^2 + \phi(r) =
\frac{1}{2} v_\infty^2 ,
\end{equation}
where we have chosen the zero of the potential $\phi(r)$ at
infinity. Conservation of angular momentum gives
\begin{equation}
\label{eq:angmomentum}
r^2 {\dot\theta} = v_0 r_0 \sin \alpha
\end{equation}
where $v_0 = \sqrt{v_\infty^2+v_{\rm esc}^2} $ is the particle speed at the Sun
location $r_0$ and $v_{\rm esc} = \sqrt{-2\phi(r_0)}$ is the escape speed at
the Sun location.
Substituting  eq.~(\ref{eq:angmomentum}) into
eq.~(\ref{eq:energy}), and defining
$u=r_0/r$, we have
\begin{equation}
\label{eq:orbit}
\frac{du}{d\theta} =
\pm \frac{1}{\sin\alpha} \sqrt{ K(u) - u^2 \sin^2\alpha } ,
\end{equation}
where the upper (positive) sign in eq.~(\ref{eq:orbit}) applies to particles
going in, and the lower (negative) sign to particles going out. Here
\begin{equation}
K(u) = \frac{ v_\infty^2 - 2 \phi(r_0/u) } {v_0^2 }
\end{equation}
is the ratio of the particle kinetic energies at galactocentric distances
$r=r_0/u$ and $r_0$.
We evaluate $K(u)$ for an NFW, a logarithmic, and a Keplerian gravitational
potential, and for the free particle case.  Then we
integrate eq.~(\ref{eq:orbit}) from $\theta=0$ to $\theta=\pi-\alpha_{\infty}$
with initial condition $u(0)=0$ to find
\begin{equation}
\label{eq:ainf}
\alpha_\infty = \cases{
  \pi + \Theta(\alpha) - \Psi(\alpha) , & for  $ \cos\alpha \ge 0 $, \cr
  \pi - \Theta(\alpha) , & for  $ \cos\alpha \le 0 $, }
\end{equation}
where
\begin{equation}
\label{eq:Theta}
\Theta(\alpha) = \sin\alpha \int_0^1 \frac{du}{\sqrt{K(u)-u^2\sin^2\alpha}},
\end{equation}
and
\begin{equation}
\Psi(\alpha) = 2 \sin\alpha \int_0^{u_\alpha}
\frac{du}{\sqrt{K(u)-u^2\sin^2\alpha}}.
\end{equation}
Here $u_\alpha$ is the solution of
\begin{equation}
K(u_\alpha) - u_\alpha^2 \sin^2 \alpha = 0;
\end{equation}
$\Psi(\alpha)$ is the angle between asymptotes for particle trajectories
passing at the Sun position with speed $v_0$.

We now present the results of calculating $\alpha_\infty$ as a function of
$\alpha$. Figure 3 plots the four functions obtained for the NFW potential, the
logarithmic potential, the Keplerian potential, and the free particle case. The
axes have been labeled in degrees.  

The graph in Figure 3 can be inverted to obtain the particle direction $\alpha$
at the Sun position once its direction at infinity $\alpha_\infty$ is given.
This inverse function is in general multi-valued: for instance, $\alpha_\infty
= 20^\circ$ in Figure 3 corresponds to two values of $\alpha$ for each
gravitational potential plotted. The physical interpretation is that for these
incoming directions of the particle stream, particles reach the Sun position
from two separate directions: one beam has passed to the ``left'' of the
galactic center, the other to the ``right''. Notice that for some stream
directions there is only one possible direction for the particle flow at the
Sun. It must be remembered that except for $\alpha_\infty = 0$ the particle
directions at the Sun position lie in the plane defined by the directions of
the galactic center and of the stream velocity at infinity.
The case $\alpha_\infty = 0$ is special, in that it corresponds to the Sun
being on the downstream axis. The values of $\alpha$ read from the graph in
Figure 3 then give the half-angle of the cone of directions from which the
particles approach the Sun.

Particles approaching the galaxy from the direction of M31 with a speed
$v_\infty=70$ km/s have $v_0=604.1$ km/s and
\begin{equation}
\alpha_\infty = 58.\!\!^\circ5,
\qquad \alpha_{\rm Kep} = \{ 113.\!\!^\circ4, 305.\!\!^\circ0 \} ,
\qquad \alpha_{\rm NFW} = 88.\!\!^\circ3,
\qquad \alpha_{\rm Log} = 81.\!\!^\circ4.
\end{equation}
There is only one particle flow for the NFW and the logarithmic potential.
In the NFW case the particle velocity at the Sun position is almost
tangential.  This is because in those trajectories that pass by the Sun, the
Sun is almost at the point of closest approach to the galactic center. At this
point, the particle radial velocity vanishes, and the particle velocity
is purely tangential.

In Figures 4 and 5, we present examples of two orbits for an NFW potential.
Figure 4 is a typical case with a relatively small deviation of the angle from
its value at infinity, while Figure 5 is close to the extreme case in which the
orbit has a purely tangential component at the location of the Sun.

\begin{figure}
\epsfig{file=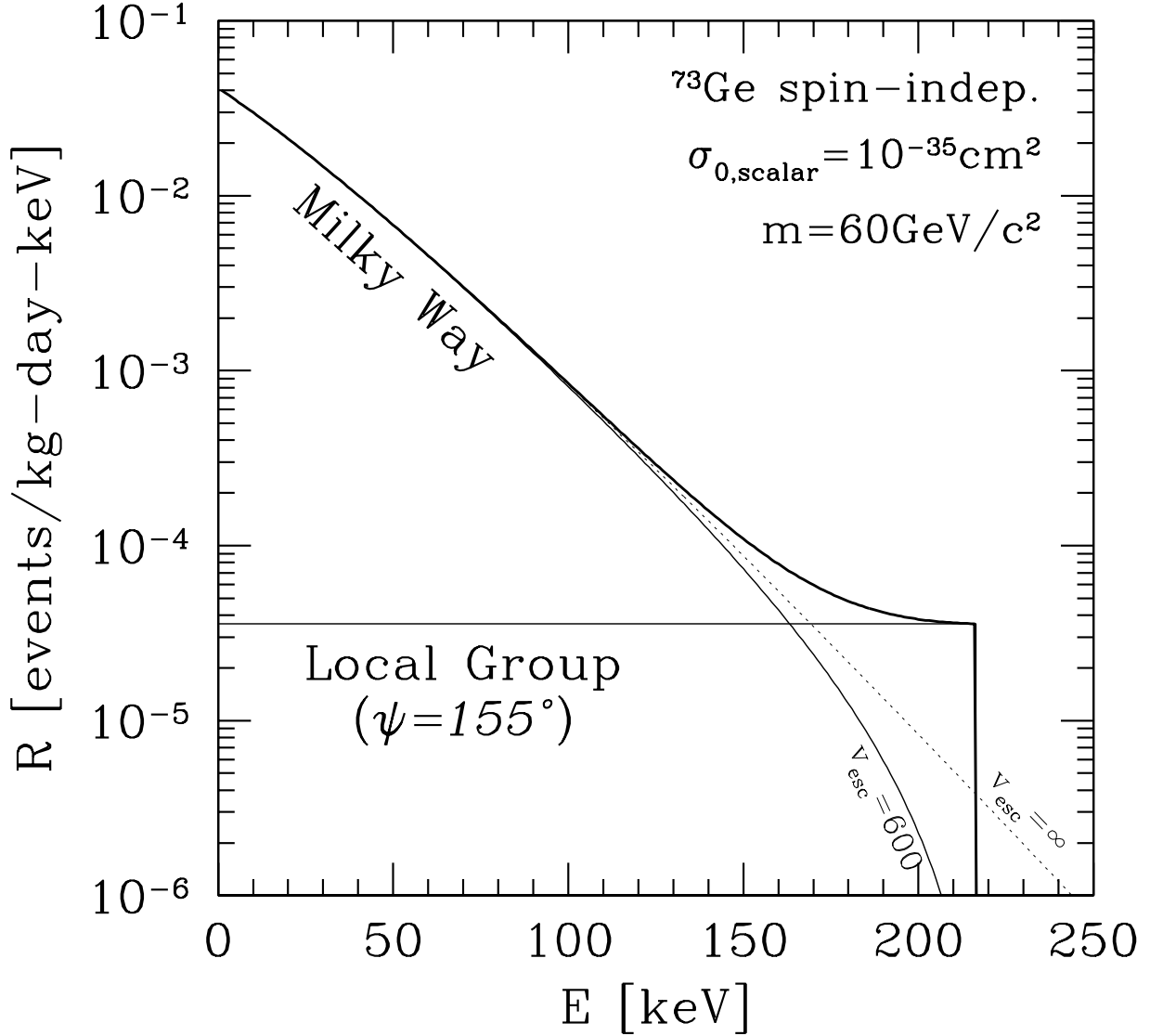,width=\textwidth}
\caption{
  Expected recoil spectra for Milky Way and Local Group WIMPs. Detector and
  WIMP parameters as indicated. The Local Group spectrum corresponds to an
  angle $\psi=155^\circ$ between the velocity of the Local Group WIMPs and the
  velocity of galactic rotation at the solar neighborhood (both velocity
  vectors starting at the location of the Earth). The Milky Way
  spectrum reflects a velocity distribution truncated at the escape speed
  (curve marked $v_{\rm esc}=600$).
The dotted line, marked $v_{\rm esc}=\infty$, does not include this
truncation. Thick lines are the sum of
  the galactic (with $v_{\rm esc}=600$ km/s) and extragalactic spectra.}
\end{figure}

\begin{figure}
\epsfig{file=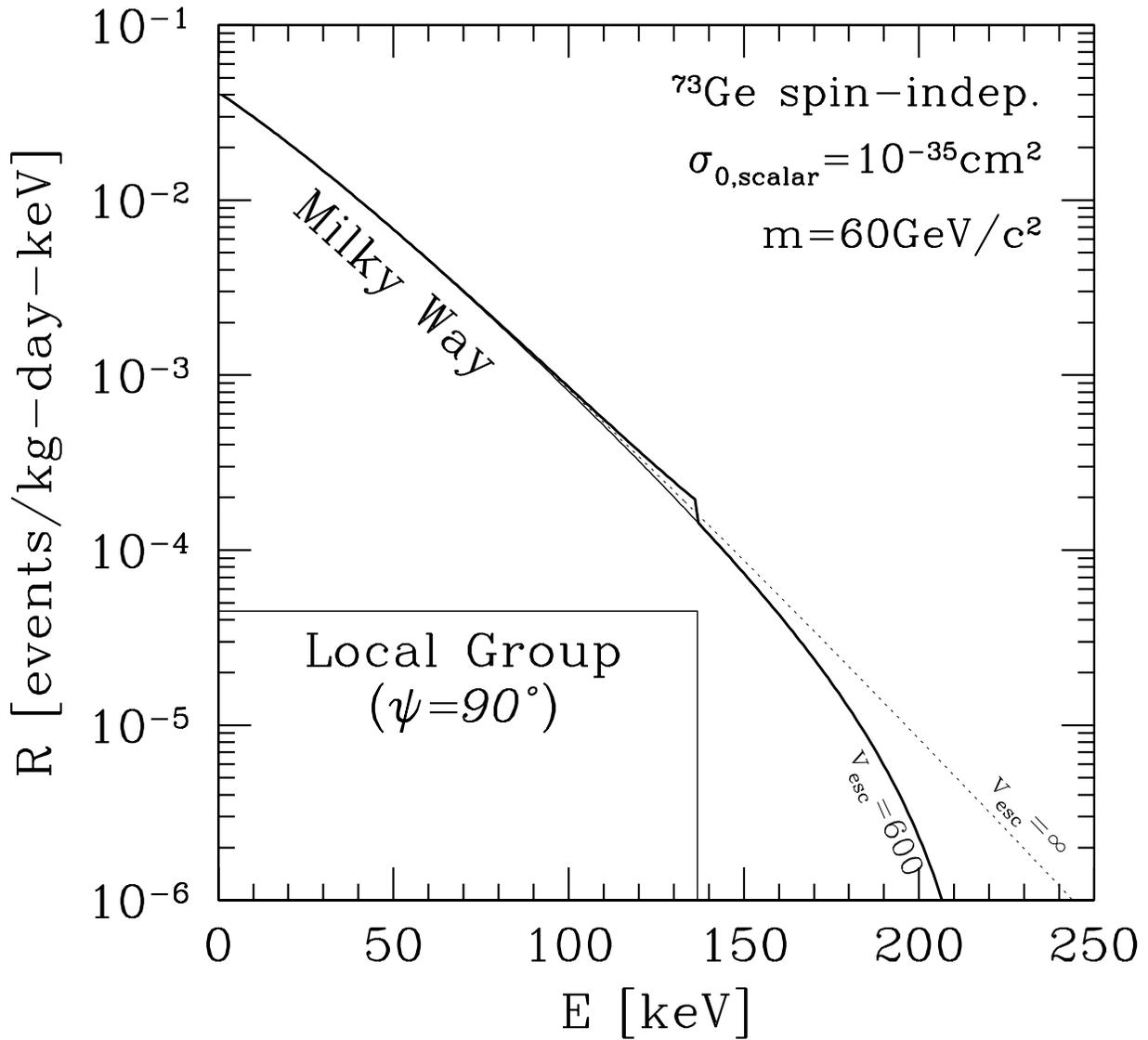,width=\textwidth}
\caption{Same as Figure 1, but for an
  angle $\psi=90^\circ$ between the velocity of the Local Group WIMPs and the
  velocity of galactic rotation at the solar neighborhood.}
\end{figure}

\begin{figure}[ht]
\epsfig{file=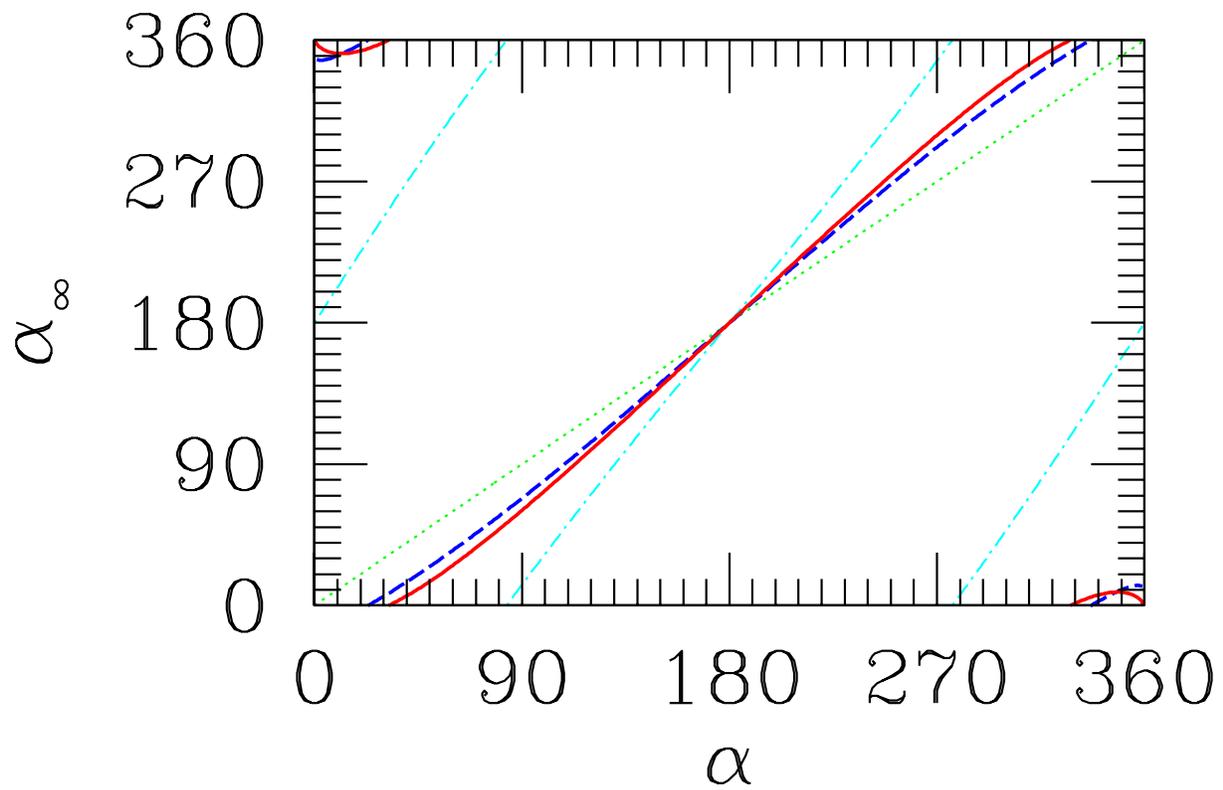,width=\textwidth}
\caption{Graph relating the particle direction at infinity $\alpha_\infty$
  to the particle direction at the Sun position $\alpha$. Clearly for
  any given incoming velocity from infinity, at most two directions in a 
  detector are possible.  Solid lines are
  for the NFW potential, dashed lines for the logarithmic potential,
  dash-dotted lines for the Keplerian potential, and dotted lines for the free
  particle case.}
\end{figure}

\begin{figure}[ht]
\begin{center}
\epsfig{file=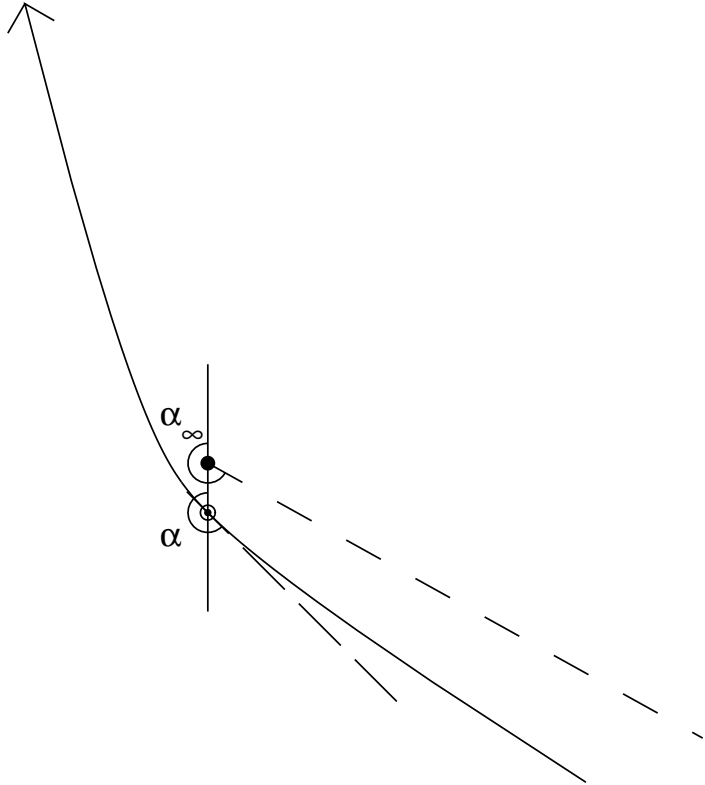,width=0.8\textwidth}
\end{center}
\caption{An example of an NFW trajectory with
  incoming angle at infinity $\alpha_\infty=225^\circ$. The lower dot indicates
  the position of the Sun, the upper dot denotes the galactic center. The
  arrival direction and the incoming direction at infinity are shown by dashed
  lines.  Also shown are the angles $\alpha_\infty$ and $\alpha$.  }
\end{figure}

\begin{figure}[ht]
\begin{center}
\epsfig{file=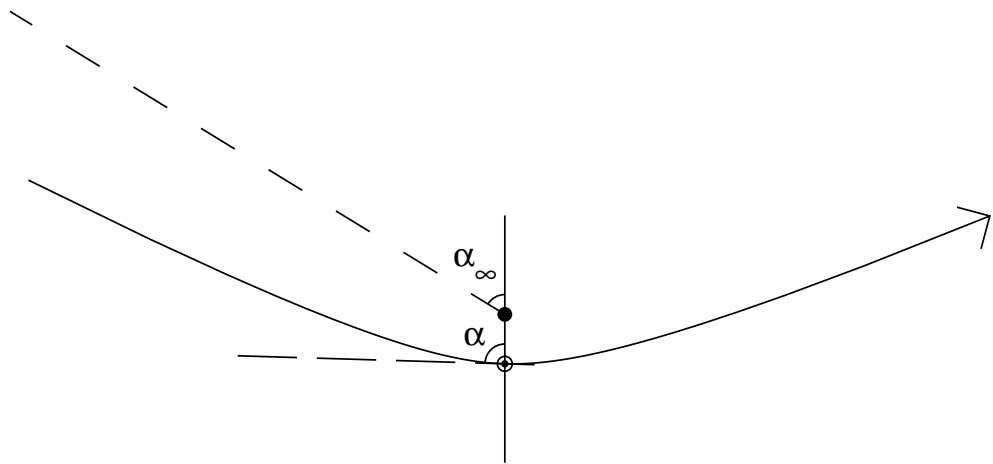,width=0.8\textwidth}
\end{center}
\caption{
  The unique NFW trajectory passing by the Sun for a particle wind
  coming from an angle of $58.\!\!^\circ5$ from the galactic center
  (i.e. from the direction of the Andromeda galaxy). Notice the almost
  tangential character of the trajectory near the Sun.}
\end{figure}

\end{document}